\documentclass[twocolumn]{article}
\usepackage{graphics,graphicx,amsmath,amssymb}
\usepackage[latin2]{inputenc}
\usepackage[magyar,english]{babel}

\newcommand{\td}[2]{\frac{\mathrm{d}{#1}}{\mathrm{d}{#2}}}
\newcommand{\z}[1]{\left({#1}\right)}
\newcommand{\sz}[1]{\left[{#1}\right]}
\newcommand{\kz}[1]{\left\{{#1}\right\}}
\newcommand{\rec}[1]{\frac{1}{#1}}
\newcommand{\m}[1]{\mathrm{#1}}

\renewcommand{\v}[1]{\mathbf{#1}}
\renewcommand{\r}[1]{(\ref{#1})}
\newcommand{\Eq}[1]{Eq.\@ (\ref{#1})}
\newcommand{\Eqs}[2]{Eqs.\@ (\ref{#1}) and (\ref{#2})}

\begin{document}
\title{Exact solutions of relativistic perfect fluid hydrodynamics for a QCD equation of state}
\author{M.~Csanád$^{1}$, M.~I.~Nagy$^{1}$, S. Lökös$^{1}$\\
$^1$Eötvös Loránd University, H-1117 Budapest, Pázmány P. s. 1/a}

\maketitle

\abstract{
We generalize a previously known class of exact analytic solutions of relativistic perfect fluid
hydrodynamics~\cite{Csorgo:2003ry} for the first time to arbitrary temperature-dependent Equation of
State. We investigate special cases of this class of solutions, in particular, we
present hydrodynamical solutions with the Equation of State determined from
lattice QCD calculations. We discuss the phenomenological relevance of these
solutions as well.
}

\section{Introduction}

The interest in relativistic hydrodynamics grew in past years mainly due to the
discovery of the almost perfect fluidity of the experimentally created Quark-Gluon-Plasma~\cite{Zajc:2007ey}.
Hydrodynamical models aim to describe the space-time picture of heavy-ion collisions
and infer the relation between experimental observables and the initial conditions.
Besides numerical simulations there is also interest in models where exact solutions
of the hydrodynamical equations are used.

In this paper we generalize a previously known class of exact solutions of relativistic
perfect fluid hydrodynamics~\cite{Csorgo:2003ry} to
the case of arbitrary, temperature dependent speed of sound. The mentioned class of
solutions form the basis of the relativistic Buda-Lund hydrodynamical model~\cite{Csanad:2003qa}. This model
yields a successful description of hadronic observables at RHIC energies (such as
the pseudorapidity and transverse momentum dependence of the azimuthal anisotropy
of different hadrons as well as the HBT radii~\cite{Csanad:2003qa}), and the reconstructed
final state in this model corresponds to simple explicit scaling solutions of
hydrodynamics. The same final state however can be achieved from many initial states, depending on the
Equation of State~\cite{Csanad:2009sk}. If one is given a temperature dependent speed of sound as Equation of State,
the solution presented in this paper thus can be used to determine
the initial state from the reconstructed final state of a heavy-ion collision. As an example, we describe the
time dependence of the system if one assumes the Equation of State from lattice QCD.

The solutions given in this paper are the first exact analytic solutions of 1+3 dimensional relativistic hydrodynamics,
to utilize an arbitrary Equation of State.\footnote{Note that it has been discussed in Ref.~\cite{Beuf:2008vd} that the
entropy flow can be calculated with an arbitrary EoS (speed of sound) from the Khalatnikov-potential, once the solution
of the general Khalatnikov equation is known.}

\section{Basic equations}

Let us adopt the following notational conventions: the fluid coordinates are $x^\mu = \z{t, \v{r}}$, where $\v{r}=\z{r_x, r_y, r_z}$ is the spatial
coordinate, and the metric tensor is $g_{\mu\nu}=diag\z{1,-1,-1,-1}$. (We denote space-time indices by Greek letters, space indices by Latin letters,
and assume the summation convention.) The fluid four-velocity is $u^\mu\equiv\gamma\z{1,\v{v}}$, where $\v{v}$ is the three-velocity and
$\gamma=1/\sqrt{1-v^2}$. The thermodynamical quantities are denoted as follows: $p$ is the pressure, $\varepsilon$ is the energy density, $\sigma$ is
the entropy density, $T$ is the temperature. If the fluid consists of individual conserved particles, or if there is some conserved charge,
then the conserved number density is denoted by $n$, and the corresponding chemical potential by $\mu$. (For more than one conserved number
densities, we may use indices to distinguish them.) All these quantities have dependence on $x^\mu$, but mostly this will not be explicitly
written out.

The basic hydrodynamical equations are the continuity and energy-momentum-conservation equations:
\begin{align}
\partial_\mu\z{n u^\mu} & = 0,\label{e:cont}\\
\partial_\nu T^{\mu \nu} & = 0\label{e:em}.
\end{align}
The energy-momentum tensor of a perfect fluid is
\begin{align}
T^{\mu\nu} =\z{\varepsilon+p}u^\mu u^\nu-pg^{\mu \nu} .
\end{align}
\Eq{e:em} can be then transformed to (by projecting it orthogonal and parallel to $u^\mu$, respectively):
\begin{align}
\z{\varepsilon+p}u^{\nu}\partial_{\nu}u^{\mu} & =\z{g^{\mu\nu}-u^{\mu}u^{\nu}}\partial_{\nu}p,\label{e:euler} \\
\z{\varepsilon+p}\partial_{\nu}u^{\nu}+u^{\nu}\partial_{\nu}\varepsilon & = 0\label{e:energy}.
\end{align}
\Eq{e:euler} is the relativistic Euler equation, while \Eq{e:energy} is the relativistic form of the energy conservation equation. In
Appendix~\ref{s:app:Teq} we recall the well-known fact that \Eq{e:energy} is equivalent to the entropy conservation equation:
\begin{align}\label{e:scont}
\partial_\mu\z{\sigma u^\mu}=0 .
\end{align}

An analytic hydrodynamical solution is a functional form of $\varepsilon$, $p$, $T$, $u^\mu$ (and, if dealt with, $n$), which solves
\Eqs{e:euler}{e:energy}, and, if present, $n$ also solves \Eq{e:cont}. The quantities $\varepsilon$, $p$, $T$, and also $\sigma$, and $n$ are
subject to the Equation of State (EoS), which closes the set of equations. We investigate the following EoS:
\begin{align}\label{e:eos}
\varepsilon = \kappa\z{T} p ,
\end{align}
and for the case when there is a conserved $n$ number density, the additional assumption is
\begin{align}\label{e:tdef}
p=nT. 
\end{align}

For the case of $\kappa\z{T}=\kappa$ constant, an ellipsoidally symmetric solution of the hydrodynamical equations is presented in Ref.~\cite{Csorgo:2003ry}:
\begin{align}\label{e:usol0}
u^\mu = \frac{x^\mu}{\tau} ,\quad \tau=\sqrt{t^2-r^2}=\sqrt{x_\mu x^\mu} ,
\end{align}
\begin{align}\label{e:tsol0}
n = n_0\frac{V_0}{V}\nu\z{s},\quad T = T_0\z{\frac{V_0}{V}}^{\rec{\kappa}}\rec{\nu\z{s}} ,
\end{align}
with $\nu\z{s}$ being an arbitrary function and 
\begin{align}\label{e:V0s}
s = \frac{r_x^2}{X^2} + \frac{r_y^2}{Y^2} + \frac{r_z^2}{Z^2},\quad V=\tau^3 ,
\end{align}
where $X$, $Y$, and $Z$ are the time ($t$) dependent principal axes of an expanding ellipsoid. They have the explicit time dependence as
\begin{align}\label{e:Z0}
X = \dot X_0 t,\quad Y = \dot Y_0 t, \quad Z = \dot Z_0 t
\end{align}
with $\dot X_0$, $\dot Y_0$, $\dot Z_0$ constants. The quantity $s$ has ellipsoidal level surfaces, and obeys $u^\nu\partial_\nu s=0$. We call $s$ a
\emph{scaling variable}, and $V$ the effective volume of a characteristic ellipsoid\footnote{
Note that with the $X\z{t}$, $Y\z{t}$, $Z\z{t}$ time-dependent axes introduced as here, we can write the velocity field as
\begin{align}
\v{v}=\z{\frac{\dot X}{X}r_x,\frac{\dot Y}{Y}r_y,\frac{\dot Z}{Z}r_z},
\end{align}
which underlines the resemblance of this solution to certain non-relativistic exact solutions with Hubble-like expansion~\cite{Csorgo:2001ru}.
}.
This solution is \emph{non-accelerating}, ie. obeys $u^\nu\partial_\nu u^\mu=0$. In the next section we present a generalization
of this class of solutions to more general EoS. The new solutions will be presented in Section~\ref{s:sols}, while
 Section~\ref{s:eoseqs} details their derivation.

\section{General Equation of State}\label{s:eoseqs}

In order to find more general solutions, where a temperature dependent EoS can be used (as in \Eq{e:eos}), for a given $u^\mu$ velocity field
we may \emph{define} the $V$ and $s$ quantities by their properties that

\begin{align}\label{e:V1}
u^\mu\partial_\mu V = V \partial_\mu u^\mu,\quad u^\mu\partial_\mu s = 0 .
\end{align}
With these quantities, \Eq{e:cont} is automatically solved (for the case when there is a conserved charge present) if
\begin{align}\label{e:nsol}
n = n_0\frac{V_0}{V}\nu\z{s} ,
\end{align}
again, with arbitrary $\nu\z{s}$ function. To solve the \r{e:energy} energy equation, we must make a distinction between two possible cases.
The first case is if we take a conserved $n$ into account, and use the EoS $\varepsilon=\kappa\z{T}p$, $p=nT$ as in \Eqs{e:eos}{e:tdef}. The
second case is when we do not consider any conserved $n$. In Appendix~\ref{s:app:Teq} we show that in both of these two cases the energy equation \Eq{e:energy} can be transformed to an equation for $T$: in the first case with conserved $n$, we have
\begin{align}\label{e:T1}
u^\mu\sz{\td{\z{\kappa T}}{T}\frac{\partial_\mu T}{T} +  \frac{\partial_\mu V}{V}} = 0 ,
\end{align}
while in the case where there is no conserved $n$, we have
\begin{align}\label{e:T2}
u^\mu\sz{\frac{\partial_\mu V}{V}+\z{\rec{\kappa+1}\td{\kappa}{T}+\frac{\kappa}{T}}\partial_\mu T}=0 .
\end{align}
Remarkably, these equations are not the same (however, we may note that in the case when $\kappa=$ const., they yield the same condition).
We call these equtions the temperature equations for the two cases. With the introduction of the $f\z{T}$ function as 
\begin{align}\label{e:fT1}
f(T)=\exp\kz{\int_{T_0}^T\z{\rec{\beta}\td{}{\beta}\sz{\kappa\z{\beta}\beta}}\m{d}\beta} 
\end{align}
for the case of conserved $n$, and as
\begin{align}\label{e:fT2}
f(T)=\exp\kz{\int_{T_0}^T\z{\frac{\kappa\z{\beta}}{\beta}+\rec{\kappa\z{\beta}+1}\td{\kappa\z{\beta}}{\beta}}\m{d}\beta} 
\end{align}
for the case of vanishing $n$, the temperature equations can be cast in the following universal form:
\begin{align}\label{e:T3}
u^\mu\sz{\frac{\partial_\mu V}{V}+\frac{\partial_\mu f(T)}{f(T)}}=0 .
\end{align}
For any given $\kappa(T)$ function we can determine $f(T)$, and write up the solution of the above equation as
\begin{align}\label{e:Tsol}
T=f^{-1}\z{\frac{V_0}{V}\xi\z{s}}
\end{align}
with arbitary $\xi\z{s}$ function. (For convenience, we may normalize $\xi\z{s}$ so that $\xi(0)=1$.)
Knowing that $u^\mu\partial_\mu s=0$, it is easy to see that this indeed solves \Eq{e:T3}. Note that if $\kappa=$ const.,
then due to \Eq{e:Tsol}:
\begin{align}\label{e:fkconst}
f(T)=\z{\frac{T}{T_0}}^{\kappa} \;\Rightarrow\;\; T = \z{\frac{V_0}{V}}^{1/\kappa}\xi(s)^{1/\kappa}.
\end{align}

As a generalization of the solution recalled in the previous section, we assume that $u^\mu$ and thus $s$ and $V$ has the same forms
as in \Eqs{e:usol0}{e:V0s}.
\begin{align}\label{e:uVs}
u^\mu=\frac{x^\mu}{\tau},\quad V=\tau^3,\quad s=\frac{r_x^2}{\dot{X}_0^2t^2} + \frac{r_y^2}{\dot{Y}_0^2t^2} + \frac{r_z^2}{\dot{Z}_0^2t^2} .
\end{align}
For this velocity field, $u^\nu\partial_\nu u^\mu=0$, so the remaining equation, the Euler equation of \r{e:euler}
is equivalent to
\begin{align}\label{e:euler0}
\partial_\mu p=u_\mu u^\nu\partial_\nu p .
\end{align}
In the case of vanishing $n$, using the thermodynamic relation $\m{d}p=\sigma\m{d}T$, \Eq{e:euler0} simplifies to
\begin{align}
\partial_\mu T=u_\mu u^\nu\partial_\nu T ,
\end{align}
and using the expression of $T$ from \Eq{e:Tsol} and the definition of $u^\mu$ and $V$ from \Eq{e:uVs}, we find that it is equivalent (for any $\kappa\z{T}$, thus for any $f(T))$ to 
\begin{align}\label{}
{f^{-1}}'\z{\frac{V_0}{V}\xi\z{s}}\frac{\xi'\z{s}}{\xi\z{s}}\partial_\mu s=0\quad\Rightarrow\quad \xi\z{s}=1.
\end{align}
In the case of non-vanishing $n$, using  \Eq{e:nsol} and $p=nT$, the Euler equation for our non-accelerating
velocity field transforms to the following equation:
\begin{align}\label{}
T\partial_\mu n+n\partial_\mu T = Tu_\mu u^\nu\partial_\nu n+nu^\mu u^\nu\partial_\nu T .
\end{align}
Substituting $n$ and $T$ from \Eqs{e:nsol}{e:Tsol}, and the definition of $V$, we get from this equation the following constraint:
\begin{align}\label{e:euln3}
\sz{\frac{\nu'\z{s}}{\nu\z{s}}+ \varphi \z{\frac{V_0}{V}\xi\z{s}} \frac{\xi'\z{s}}{\xi\z{s}}} \partial_\mu s = 0 ,
\end{align}
where we have introduced the following function:
\begin{align}
\varphi(y) = \frac{y{f^{-1}}'(y)}{f^{-1}(y)}
\end{align}
Since $\partial_\mu s\neq 0$, we see from \Eq{e:euln3} that there are two cases: for any EoS (ie.\ for any $\kappa\z{T}$ and thus any $\varphi(T)$ function)
we get a solution if $\nu\z{s}=\xi\z{s}=1$. The other possibility is if $\kappa=const$. It is easy to see that this case is
equivalent to $\varphi=\kappa^{-1}=const$, and so \Eq{e:euln3} is solved if $\xi = \nu^{-1/\kappa}$ and so from
\Eq{e:fkconst} we get $T=T_0\z{V_0/V}^{1/\kappa}\nu^{-1}(s)$, i.e. the same as in \Eq{e:tsol0}. In this case
we indeed obtain the known solution of Ref.~\cite{Csorgo:2003ry}, recited in Eqs.~\r{e:usol0}--\r{e:V0s}.

\section{New solutions for general Equation of State}\label{s:sols}

Summarizing and rewriting the results presented in the previous section, we found new solutions to the relativistic hydrodynamical equations for arbitary
$\varepsilon=\kappa\z{T}p$ Equation of State, and these are the first solutions of their kind (i.e. with a non-constant EoS). In the case where we do not consider any conserved $n$ density, the solution can be presented in the following
form, in terms of $u^\mu$, $\sigma$ and $T$, with $T$ given in an implicit form:
\begin{align}
\sigma &= \sigma_0 \frac{\tau_0^3}{\tau^3} ,\label{e:Tsol:s:0}\\
u^\mu & = \frac{x^\mu}{\tau} ,\\
\frac{\tau_0^3}{\tau^3} & = 
\exp\kz{\int_{T_0}^T\z{\frac{\kappa\z{\beta}}{\beta}+\rec{\kappa\z{\beta}+1}\td{\kappa\z{\beta}}{\beta}}\m{d}\beta} .
\label{e:Tsol:s}
\end{align}
Also, for the case when the pressure is expressed as $p=nT$ with some conserved $n$ density, the new solution is written in terms of $u^\mu$, $T$ and $n$ as
\begin{align}
n &= n_0 \frac{\tau_0^3}{\tau^3} ,\\
u^\mu & = \frac{x^\mu}{\tau} ,\\
\frac{\tau_0^3}{\tau^3} & =
\exp\kz{\int_{T_0}^T\z{\rec{\beta}\td{}{\beta}\sz{\kappa\z{\beta}\beta}}\m{d}\beta} . \label{e:Tsol:nT}
\end{align}
Note that these solutions form simple generalization of the $\nu\z{s}=1$ case of the solutions of Ref.~\cite{Csorgo:2003ry}.
Also note that in the case when $p=nT$ and $n$ is conserved, for some choices of the $\kappa\z{T}$ function our solution becomes ill-defined.
The criterion that $\td{}{T} \z{\kappa\z{T}T}$ should be positive limits the applicability of solutions for the case of conserved $n$ presented
here. In the case when for some $T$ range $\td{}{T} \z{\kappa\z{T}T}$ becomes negative, the implicit form of \Eq{e:Tsol:nT} cannot be inverted
to give a unique $T\z{\tau}$ function. Such domains of $T$ indeed might exist in some parametrizations of the lattice QCD Equation of State around the
quark-hadron transition temperature (as detailed in the next section, in particularly on Fig.~\ref{f:validity}). However, even for these cases, one can use the solution without conserved $n$ presented here as a physically relevant
solution, since at the transition temperature a conserved density $n$ yielding pressure as $p=nT$ cannot be identified.

Let us briefly mention another possibility, when $\kappa$ is a function of the pressure $p$ and not that of the temperature $T$.
In this case a solution can be written up, similarly to the previous ones as
\begin{align}
\sigma &= \sigma_0 \frac{\tau_0^3}{\tau^3} ,\label{e:psol:s}\\
u^\mu & = \frac{x^\mu}{\tau} ,\\
\frac{\tau_0^3}{\tau^3} & = 
\exp\kz{\int_{p_0}^p\z{\frac{\kappa\z{\beta}}{\beta}+\td{\kappa\z{\beta}}{\beta}}\frac{\m{d}\beta}{\kappa\z{\beta}+1}} ,
\label{e:psol:p}
\end{align}
i.e. almost the same as in \Eq{e:Tsol:s}, except that here the integration variable is the pressure $p$. If however the pressure
can be written as a function of temperature, i.e. as $p(T)$, an integral-transformation can be made with and we get back
\Eq{e:Tsol:s}, so in this case these solutions are identical. This solution may be used if a $\kappa(p)$ function is given 
(without relation to the temperature) by an arbitrary energy density function $\varepsilon(p)=\kappa(p)p$.

\section{Application}
\begin{figure}
 \begin{center}
 \includegraphics[angle=270,width=0.49\textwidth]{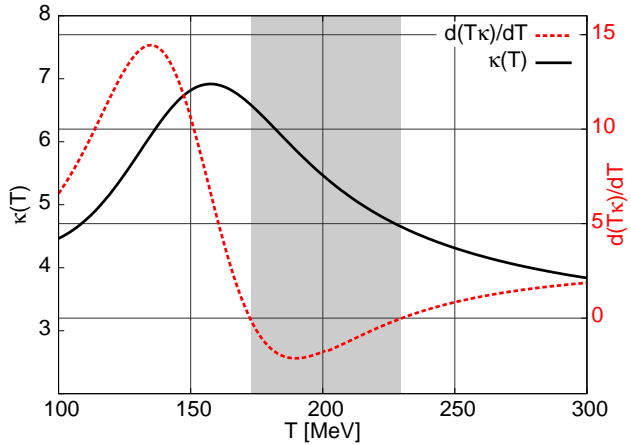}
 \end{center}
 \caption{The temperature dependence of the EoS parameter $\kappa$ from  Ref.~\cite{Borsanyi:2010cj} is shown with the solid
black curve. Note that in the shaded $T$ range (173 MeV - 230 MeV) $\td{}{T} \z{\kappa\z{T}T}$ (red dashed line) becomes
negative, thus the implicit form of \Eq{e:Tsol:nT} cannot be inverted to give a unique $T\z{\tau}$ function. Hence we will substitute this $\kappa(T)$ in the hydrodynamic solution shown in Eqs.~\r{e:Tsol:s:0}--\r{e:Tsol:s}.}\label{f:validity}
\end{figure}

Recently a QCD equation of state has been calculated by the Budapest-Wuppertal group in Ref.~\cite{Borsanyi:2010cj}. Here
(in their Eq.~(3.1) and Table 2) they give a parametrization of the trace anomaly as a function of temperature. Hence the pressure, the energy density and finallly the EoS parameter $\kappa$ can be calculated, as a function of the temperature. We did this
calculation, and got the $\kappa(T)$ function as shown in Fig.~\ref{f:validity}. Note however, that in this calculation for some $T$ range $\td{}{T} \z{\kappa\z{T}T}$ becomes negative, as also shown in Fig.~\ref{f:validity}. Hence the implicit form of \Eq{e:Tsol:nT} cannot be inverted to give a unique $T\z{\tau}$ function. We can still use the solution without conserved number density $n$, presented in Eqs.~\r{e:Tsol:s:0}--\r{e:Tsol:s}.

\begin{figure}
 \begin{center}
 \includegraphics[angle=270,width=0.49\textwidth]{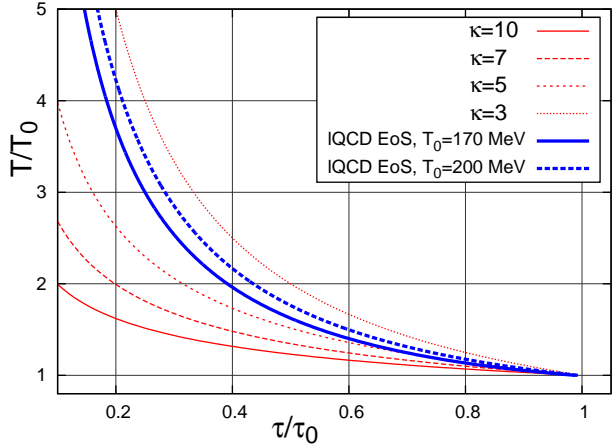}
 \end{center}
 \caption{Time dependence of the temperature $T(\tau)$ (normalized with the freeze-out time $\tau_0$
and the freeze-out temperature $T_0$) is shown here. The four thin red lines show this dependence in
case of constant $\kappa$ values, while the thicker blue lines show results based on the EoS of Ref.~\cite{Borsanyi:2010cj}.
The resulting curve slightly depends on the value of $T_0$. It is clear however, that the temperature fall-off is almost as fast
in the QCD EoS case as in the case of fixed $\kappa=3$, which resembles a relativistic ideal gas. This means, that a fixed
freeze-out temperature (which cannot vary too much due to the known quark-hadron transition temperature) results in a very high initial temperature.}\label{f:ttau}
\end{figure}

We utilized the obtained $\kappa(T)$ and calculated the time evolution of the temperature of the fireball from this
solution of relativistic hydrodynamics. The result is shown in Fig.~\ref{f:ttau}. Clearly, temperature falls off almost as
fast as in case of a constant $\kappa=3$, an ideal relativistic gas. Hence a given freeze-out temperature yields a
significantly higher initial temperature than a higher $\kappa$ (i.e. a low $c_s^2$) would.
If we fix the freeze-out temperature to 170 MeV for example,
then already at 30\% of the freeze-out time $\tau_0$ (the value of which does not affect our results) 2.5-3$\times$
higher temperatures than at the freeze-out. To give a concrete example, if $\tau_0 = 8$ fm$/c$, and $\tau_{\rm init} = 1.5$
 fm$/c$, then for $T_0 = 170$ MeV we get $T_{\rm init} \approx 550$ MeV (and even higher if $\tau_{\rm init}$ is smaller).
The QCD equation of state of Ref.~\cite{Borsanyi:2010cj} and this hydro solution yields a general $T(\tau)$ dependence.
If the freeze-out temperature $T_0$ and the time evolution duration $\tau_0 / \tau_{\rm init}$ are known, the initial
temperature of the fireball can be easily calculated.

\section{Conclusion}

We have presented the first analytic solutions of the equations of relativistic perfect fluid hydrodynamics for general temperature dependent speed of sound (ie.\ general Equation of State).
 They can be seen as generalizations of previously known exact solutions~\cite{Csorgo:2003ry}.
However, our new solutions are spherically symmetric, thus possible generalizations of them definitely are worth exploring: solutions for the
non-accelerating case and for more general ellipsoidal symmetry would be able to analytically explore the time evolution of other hadronic
observables such as the elliptic flow ($v_2$).

We have shown how to use our solutions to fully utilize the lattice QCD Equation of State for exploring the initial state of heavy-ion reactions based
on the reconstructed final state in the Buda-Lund hydrodynamical model. In $\sqrt{s_{NN}}=200$ GeV Au+Au collisions, our investigations reveal a very
high initial temperature consistent with calculations based on the measurement spectrum of low momentum direct photons~\cite{Adare:2008fqa}. If given a temperature-dependent direct photon emission function, then this
model can be used to calculate direct photon spectra to be compared to measurements, as in Ref.~\cite{Csanad:2011jq},
but with a realistic Equation of State.

\section*{Acknowledgments}
This work was supported by the  NK-101438 OTKA grant and the Bolyai Scholarship (Hungarian Academy of Sciences) of M. Csan\'ad.  The authors also would like to thank T. Cs\"org\H{o} for motivating and valuable discussions.

\appendix
\section{The entropy and the temperature equations}\label{s:app:Teq}

The fundamental thermodynamical relations connecting $\varepsilon$, $T$, $\sigma$, $p$, and any types of $n_i$ conserved charges and corresponding
$\mu_i$ chemical potentials are
\begin{align}
\varepsilon+p    & = T\sigma+\sum_i n_i\mu_i ,\label{e:thermo1}\\
\m{d}\varepsilon & = T\m{d}\sigma+\sum_i\mu_i\m{d}n_i ,\label{e:thermo2}\\
\m{d}p           & = \sigma\m{d}T+\sum_in_i\m{d}\mu_i .\label{e:thermo3}
\end{align}
In the case when there are no conserved charges, similar relations hold with all $n_i$ and $\mu_i$ variables omitted. Substituting these in the \r{e:energy}
energy conservation equation, we immediately obtain the continuity equation for the entropy density $\sigma$:
\begin{align}
T\sigma\partial_\mu u^\mu + Tu^\mu\partial_\mu \sigma +\sum_i\mu_i\z{n_i\partial_\mu u^\mu+u^\mu\partial_\mu n_i}= 0 ,
\end{align}
which is, for conserved (or vanishing) $n_i$s, equivalent to
\begin{align}
\partial_\nu\z{\sigma u^\nu} = 0 ,
\end{align}
which is the entropy conservation, \Eq{e:scont}.

In the case when there is no conserved $n$, we can substitute the \r{e:thermo2} and \r{e:thermo3} thermodynamic relations for vanishing
$n$ in \Eq{e:energy}. Using the Eos as $\varepsilon=\kappa\z{T}p$, and $\varepsilon+p =T\sigma$ and $\m{d}p=\sigma\m{d}T$ we obtain from
\Eq{e:energy} the following:
\begin{align}
T\sigma\sz{\partial_\mu u^\mu+\rec{\kappa+1}\td{\kappa}{T}u^\mu\partial_\mu T}+\kappa\sigma\ u^\mu\partial_\mu T = 0 ,
\end{align}
which is, by using \Eq{e:V1} again, equivalent to \Eq{e:T2}, as was to be demonstrated.

Next, we would like to obtain an equation for the temperature with our specific Equation of State as in \Eq{e:eos} ($\varepsilon=\kappa\z{T}p$), in the case when
there is a conserved $n$ and $p=nT$. We can substitute these into \Eq{e:energy}, and use the \r{e:cont} continuity equation for $n$ to infer that
\Eq{e:energy} is equivalent to the following:
\begin{align}
T \partial_\mu u^\mu + \td{}{T}\z{\kappa T}u^\mu \partial_\mu T = 0 .
\end{align}
Introducing $V$ by using \Eq{e:V1}, we immediately see that this is equivalent to \Eq{e:T1} , as stated in the text.

Finally, let us show how the solution for a given $\kappa(p)$, described in Eqs.~\r{e:psol:s}--\r{e:psol:p} can be obtained. 
In that case, instead of substituting the temerature to \Eq{e:energy}, we write up the equation using the $\kappa(p)$
function and the relation $\varepsilon = \kappa\cdot p$, similarly to the previous cases:
\begin{align}
u^\nu   \sz{\frac{\partial_\nu V}{V}+\z{\frac{\kappa}{p}+\frac{d\kappa}{dp}}\frac{\partial_\nu p}{\kappa+1}}=0.
\end{align}
This equation is then solved by the implicit formula on the pressure, given in \Eq{e:psol:p}

\bibliographystyle{h-physrev}
\bibliography{../../master}

\end{document}